\documentclass[letterpaper,english]{article}
\usepackage[T1]{fontenc}
\usepackage[latin1]{inputenc}
\usepackage{amsmath}
\usepackage{amssymb}

\makeatletter

\usepackage{graphicx}
\usepackage{amssymb}

\usepackage{amssymb}

\usepackage{amssymb}

\usepackage{graphicx}

\usepackage{geometry}

\usepackage{babel}

\usepackage{babel}

\usepackage{babel}

\usepackage{babel}
\makeatother
\begin{document}

\title{\textbf{Gaussian Noise Effects on the Evolution of Wealth in a Closed System of n-Economies}}

\author{J. M. Pellón-Díaz,  A. Aragonés-Muñoz
\\A. Sandoval-Villalbazo, A. Díaz-Reynoso
\\
\\Departamento de Física y Matemáticas,
\\Universidad Iberoamericana,
\\Prolongación Paseo de la Reforma 880,
\\México D. F. 01210, México.
\\ E-mail: alfredo.sandoval@uia.mx}

\maketitle

\begin{abstract}
Based on the stochastic model proposed by Patriarca-Kaski-Chakraborti that describes the
exchange of wealth between $n$ economic agents, we analyze the evolution of the corresponding
economies under the assumption of a Gaussian background, modeling the exchange parameter $\epsilon$.We demonstrate, that within Gaussian noise, the variance of the resulting wealth distribution will significantly decrease, and the equilibrium state is reached faster than in the case of a uniform distributed $\epsilon$ parameter. Also, we show that the system with Gaussian noise strongly resembles a deterministic system which is solved by means of a Z-Transform based technique.
\end{abstract}

\section{Introduction}
Mathematics can be viewed as a language used to describe our environment in order to understand it.
Economics has been trying to provide tools that may explain the evolution of wealth. However, the
different approaches used to describe this type of evolution have not yet achieved their final goal.
Econophysics provides new tools to achieve this goal \cite{1}.

Brownian motion has been used to describe financial markets for decades. The work of Robert C.
Merton and Paul A. Samuelson \cite{2} provided useful information regarding the stock markets. Most of these models are based on Brownian motion theory, describing fluctuations, in this case earnings or losses of wealth. The purpose of this work is to propose an improvement to the Patriarca-Kaski-Chakraborti model, using Gaussian noise instead of uniform noise, since in principle it appears more frequently in random processes. The Z-Transform is an analytical tool, that helps to provide information regarding the equilibrium state of each economy.

Section 2 is dedicated to a brief review of the Patriarca-Kaski-Chakraborti model. In section
3, we incorporate Gaussian noise in the $\epsilon$ parameter to the model, and we compare the effects of
uniform noise and Gaussian noise in the evolution of economies. In section 4, we solve a deterministic system using the Z-Transform technique, in order to compare this type of evolution with its counterpart in the presence of Gaussian noise. Finally, section 5 is dedicated to final remarks.

\section{Equations of the Patriarca-Kaski-Chakraborti Model Generalized to n-Economies}

To model a closed economy of n-agents interacting with each other, some laws must rule the system
(as in physics the conservation of energy holds). For transactions in a closed system, the conservation
of wealth is assumed to hold, since the exchanges only involve the agents.

\begin{equation}
x_i + x_j +...+x_n =x'_i +x'_j+.....x'_n,
\end{equation}
where $x_n$ is the wealth of the $n$th-economy before a given transaction and $x'_n$ is the wealth after the transaction.

The basic law describing the evolution of the set of economies is given by the following expression \cite{3} \cite{4}.

\begin{equation}
x_{i,j}=\lambda_j x_{i-1,j}+\epsilon_j \sum_{j=1}^{n}((1-\lambda_j)x_{i-1,j}),
\end{equation}
where $x_{i,j}$ is the wealth for the $j$th-agent in the $i$th-transaction, $\lambda_j$ is the fraction of wealth saved for the $j$th-agent,$\epsilon_j$ is a random number between 0 and 1 uniformly distributed for the $j$th-agent and $n$ is the number of economies. $\epsilon_j$ corresponds to a probability, so that $\sum_{j=1}^{n} \epsilon_j=1$.

We propose here, that the probability of the $n$th-economy is to be obtained from a vector of random numbers given by $V_{\lambda}=\left[u_1,u_2,...,u_j\right]$ using a director cosine:

\begin{equation}
\epsilon_j=\left(\frac{u_j}{\sqrt{V_{\lambda}\cdot V_{\lambda}}}\right)^2.
\end{equation}

For $n$-economies there will exist $_nC_j$ values of the change of wealth in a given transaction $\Delta x$. We define them by:

\begin{equation}
\Delta x_{i,(a,b)}=\epsilon_b (1-\lambda_a)x_a-\epsilon_a (1-\lambda_b)x_b,
\end{equation}

where $a$ and $b$ are related to $j$ number of agents (which means the difference of wealth between two agents $a$ and $b$).

The distribution function that is adjusted by means of numerical simulations, is a Gamma distribution function \cite{3}. In the next section, we will establish a counterpart of the Gamma distribution using a Gaussian background in order to describe the noise produced during the transactions, for the n-economies model.

\section{Gaussian Noise in the Closed n-Economies Model}

The need to approach in a more precise way to Brownian motion processes has lead us to modify
the background model. It is proposed to describe the exchange of wealth parameter "$\epsilon$" of Eq. 2 with a Gaussian distribution f(x), which represents the probability of wining or losing a certain
amount in the transaction.

\begin{equation}
f(x)=\frac{1}{\sqrt{2 \pi}\sigma}e^{\frac{-(x-<x>)^2}{2 \sigma^2 }}
\end{equation}

Here, the distribution has a mean given by $<x>=1/2$ and standard deviation $\frac{1}{12}$, so that the area under the curve within $[\frac{1}{2}-6\sigma, \frac{1}{2}+6\sigma]$ is equal to $1$. Of course, this curve may be sharper than the one here proposed. Fig. 1 shows a comparison between a Gaussian distribution, using the parameters mentioned before, and a uniform distribution.

\begin{figure}

\begin{centering}
\includegraphics[
  width=3.5in,
  height=2.5in]{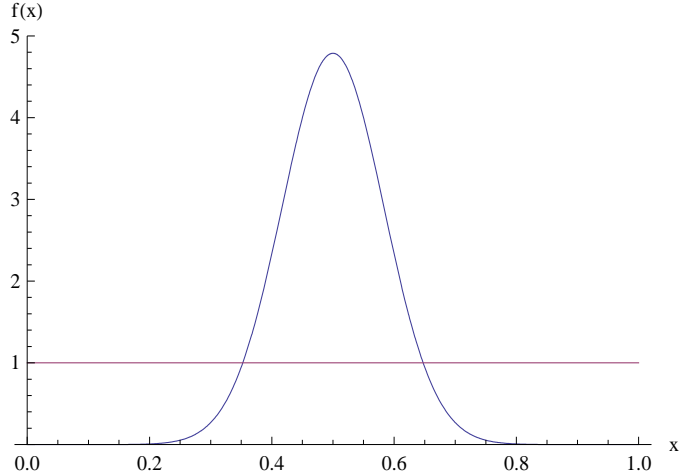}

\end{centering}

\caption{Two possibilities for modeling noise for the wealth parameter $\epsilon$ present in Eq. 2. }

\end{figure}

The corresponding probability  given in an interval $a$ to $b$ reads:

\begin{equation}
\int_{a}^{b}\frac{1}{\sqrt{2 \pi}\sigma}e^{\frac{-(x-<x>)^2}{2\sigma^2 }}dx.
\end{equation}

 Figure 2 shows the description of the wealth with both backgrounds. According to the value of
the saving parameter $\lambda$, in this case equal to 0.9, the distribution looks taller and thinner than the one corresponding to uniform noise, which means that the wealth doesnÕt vary as much as in the uniform noise case (the possibility of losing a large amount of money decreases).
 
The amount of wealth exchanged on a series of transactions depends on different factors as well
as its final distribution.

The initial amount of money does represent a fact when considering short term transactions,
but it does not represent a significant fact in long term transactions or in the equilibrium state.

An important fact to consider, is the fraction of wealth saved for each economic agent, $\lambda_j$ on the series of transactions. As this parameter increases the variance for wealth distribution decreases, leading to equilibrium faster than in the uniform noise case, as shown in Fig. (3). Also, the variance reduction between the Gaussian distribution vs. the uniform distribution is about 32$\%$.

\begin{figure}

\begin{centering}
\includegraphics[
  width=5.5in,
  height=3.5in]{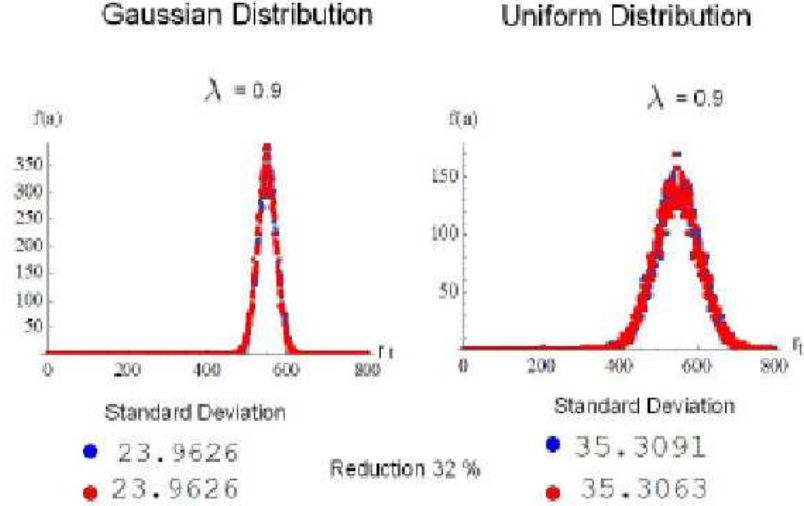}

\end{centering}

\caption{Comparison of Gaussian distribution vs. Uniform Distribution}

\end{figure}

The difference on the way the economic agents exchange money yields significant changes in
the evolution of the economic systems. As a result of it, their equilibrium states vary. An uniformdistribution means that it is equally probable to win or lose different percentage of wealth on a single transaction, meanwhile a Gaussian distribution means that it is more likely to win or lose
small percentages of wealth, and the probability of wining or losing large amounts decreases as the number of transactions increases.

\section{Z-Transform Approach}

It is important to compare the results previously shown with a suitable analytical approach. The
Z-Transform provides such an analytical tool and allows to obtain valuable information about the
evolution of the economies. This transform converts a discrete time domain set of equation into
a complex frequency domain representation, where the resulting system of algebraic equations can
be solved. To complete this idea, we wish to remark that when the noise of the system is narrowed
as in the case of a Dirac Delta function, the problem will become deterministic.

We recall that the Z-Transform is defined as

\begin{equation}
X(z)=Z(x(m))=\sum_{m=0}^{\infty} x(m) z^{-m},
\end{equation}
where $x(m)$ represents the amount of money in a given transaction $m$.

Applying the Z-Transform to Eq. (2) for constant $\epsilon$, the system will become algebraic. The corresponding matrix $S$ for the system of two economies is given by:

\begin{equation}
S=\left(
         \begin{array}{cc}
           z-\epsilon(1-\lambda_x)-\lambda_x & \epsilon (-1+\lambda_y) \\
           (1-\epsilon)(-1+\lambda_x) & z-(1-\epsilon)(1-\lambda_y)-\lambda_y \\
         \end{array}
       \right).
\end{equation}

The arrays corresponding to mx and my for each economy read as

\begin{equation}
m_x=\left(
      \begin{array}{cc}
        x_0 z & \epsilon(-1+\lambda_y) \\
        y_0 z & z-(1-\epsilon)(1-\lambda_y)-\lambda_y \\
      \end{array}
    \right),
\end{equation}

\begin{equation}
m_y=\left(
      \begin{array}{cc}
        z-\epsilon(1-\lambda_x)-\lambda_x & x_0 z \\
        (1-\epsilon)(-1+\lambda_x) & y_0 z \\
      \end{array}
    \right).
\end{equation}

The determinant of the system is easily calculated and the result is

\begin{equation}
Det(S)=z^2+\lambda_x-\epsilon \lambda_x +\epsilon \lambda_y + z(-1-\lambda_x+\epsilon \lambda_x-\epsilon \lambda_y).
\end{equation}

In order to analyze the stability properties of the system, we equal this result to zero and solve
the resulting equation. One of the roots will always be $z=1$, which means that the system will tend to a constant value. The other root is $z=\lambda_x-\epsilon \lambda_x + \epsilon \lambda_y$, and if $\lambda_x=\lambda_y$ then $z=\lambda_x$.

The next step is to obtain the inverse Z-Transform of the quotient of the determinant of $\frac{det(m_x)}{det(s)}$ and $\frac{det(m_y)}{det(s)}$. In order to illustrate this analytical result we show an example in which we arbitrarily assign the numerical values $\lambda_x=0.95$, $\lambda_y=0.8$, $x_0=1000$, $y_0=2000$ and $\epsilon=0.51$ (Fig. (3) and Fig. (4)).

\begin{figure}

\begin{centering}
\includegraphics[
  width=3.5in,
  height=2.5in]{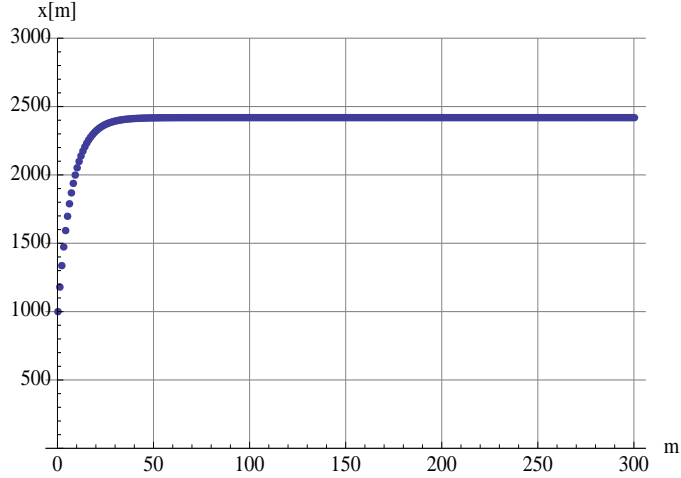}

\end{centering}

\caption{The $x(m)$ economy will win the money, and tends to a constant value (equilibrium).}

\end{figure}
\begin{figure}

\begin{centering}
\includegraphics[
  width=3.5in,
  height=2.5in]{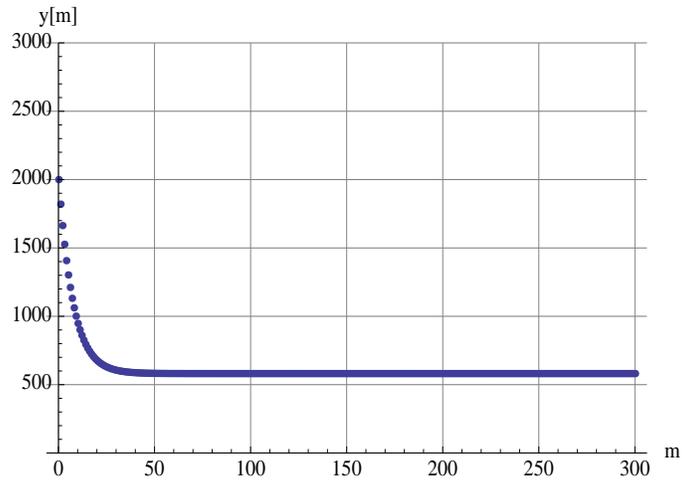}

\end{centering}

\caption{The $y(m)$ economy loses money and also tends to equilibrium}

\end{figure}

The comparison of the results obtained from the numerical simulation, with those obtained from
the deterministic model shows an interesting concordance.

\begin{figure}

\begin{centering}
\includegraphics[
  width=3.5in,
  height=2.5in]{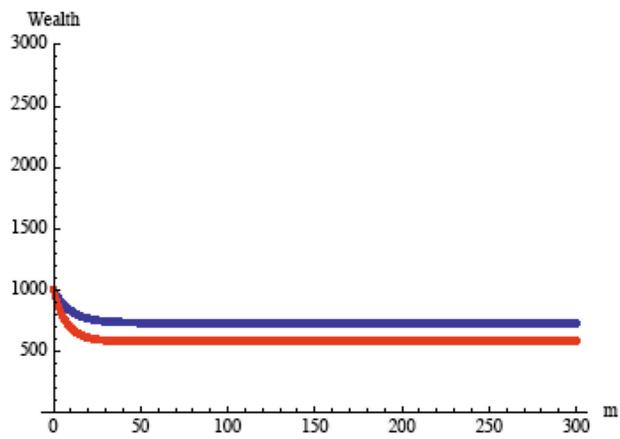}

\end{centering}

\caption{Comparison of the results between the numerical simulation and the deterministic model.
The upper line represents the numerical simulation. Both tend to equilibrium.}

\end{figure}

\section{Final remarks}

The use of a Gaussian distribution allows to show that the range of winning and losing in a given
transaction is significantly reduced, this fact leads to a more controlled model. This means that with
this kind of noise the system tends to equilibrium faster. Also for modeling the noise, the exchange
parameter is as important as the saving parameter in order to reduce the variance in the wealth
distribution between economic agents. The Z-Transform technique allows to qualitatively reproduce
the results obtained from the numerical simulation (Fig. (5)), making us capable to resemble the
evolution of a stochastic system with its deterministic counterpart. Gaussian noise can be viewed
as a background in which the risk of losing large amounts of money in each transaction is strongly
decreased . Further ideas regarding the roll of noise in the exchange parameter will be addressed
in future work.

\end{document}